\documentclass[11pt,a4paper]{llncs}
\usepackage{amsmath}
\usepackage{amssymb}
\usepackage{graphicx}
\usepackage{marvosym}
\usepackage{url}
\usepackage{fancyhdr}
\usepackage{times}
\usepackage{epsfig}
\usepackage{graphicx}
\usepackage{amsmath}
\usepackage{booktabs}
\usepackage{siunitx} 
\usepackage{amssymb}
\usepackage{graphicx}
\usepackage{wrapfig}
\usepackage{lscape}
\usepackage{rotating}
\usepackage{epstopdf}
\usepackage{graphicx} 
\usepackage{wrapfig}
\usepackage[utf8]{inputenc}
\usepackage[english]{babel}
\usepackage{url}
\usepackage{multirow}
\usepackage{hyperref}
\usepackage{float}
\usepackage{adjustbox}

\usepackage{subcaption,graphicx}
\usepackage{geometry}
\usepackage{xurl}
\newcommand{\keywords}[1]{\par\addvspace\baselineskip
\noindent\keywordname\enspace\ignorespaces#1}

\fancypagestyle{firstpage}{\fancyhf{}
}

\begin{document}


\title{\LARGE{Deep learning pipeline for image classification on mobile phones}}


%
%
\author{\large{Muhammad Muneeb  \and Samuel F. Feng \and Andreas Henschel}}
\institute{\large{Department of Mathematics and Department of Electrical Engineering and Computer Science, Khalifa University of Science and Technology,\\ Abu Dhabi, UAE}}

%


%
%


\maketitle

\thispagestyle{firstpage}

\begin{abstract}
This article proposes and documents a machine-learning framework and tutorial for classifying images using mobile phones. Compared to computers, the performance of deep learning model performance degrades when deployed on a mobile phone and requires a systematic approach to find a model that performs optimally on both computers and mobile phones. By following the proposed pipeline, which consists of various computational tools, simple procedural recipes, and technical considerations, one can bring the power of deep learning medical image classification to mobile devices, potentially unlocking new domains of applications. The pipeline is demonstrated on four different publicly available datasets: COVID X-rays, COVID CT scans, leaves, and colorectal cancer. We used two application development frameworks: TensorFlow Lite (real-time testing) and Flutter (digital image testing) to test the proposed pipeline. We found that transferring deep learning models to a mobile phone is limited by hardware and classification accuracy drops. To address this issue, we proposed this pipeline to find an optimized model for mobile phones. Finally, we discuss additional applications and computational concerns related to deploying deep-learning models on phones, including real-time analysis and image preprocessing. We believe the associated documentation and code can help physicians and medical experts develop medical image classification applications for distribution.
\end{abstract}


\section{Introduction}
\label{sec:introduction}
Disease diagnosis plays a crucial role in healthcare, and for many conditions, medical image data can aid in diagnosing diseases \cite{Mitra2015,DarshiniVelusamy2014,Bauer2013,Lin2018,Islam2018,Bhattacharya2021,Liang2021,Soomro2021}. Recent research has made these diagnostic tools more efficient, primarily through deep-learning methods \cite{Litjens2017}. However, these methods typically require relatively powerful and expensive computational hardware (e.g. modern GPUs), which may not be available in remote or poor areas of the world that lack modern infrastructure. This study proposes a pipeline for classifying images using mobile phones with a remote computer for model training. Even though a central server is required for training, there are free options \cite{googlecolab,azurenotebook} which are sufficient for training the model.

Due to the widespread availability of mobile phones and applications, tasks such as image classification can now be completed at the point-of-care. The deployment of medical image classification models on mobile phones can lead to several technical issues. For example, the model's performance when trained or tested on a computer degrades significantly when the same model is deployed on a mobile phone. Neural network models easily become large enough for phones' memory, and images captured in real time can degrade classification performance. Furthermore, it is unclear how to build a proper workflow connecting training data, often from different sources with varying image sizes and quality, with a model deployed on a smartphone to aid diagnosis. We highlight these issues (and other) and provide feasible solutions to tackle them. The result is an amalgamation of best practices taken from the modern deep learning and data science landscape, including implementations for parameter reduction (Section \ref{subsection:parameterreduction}) and data augmentation (Section \ref{subsubsection:dataaugmention}), resulting in a cost-effective diagnosis pipeline that can be used for medical image classification aiding expert in the diagnosis of the diseases.

The following section explains the differences and similarities between similar projects and the pipeline proposed in this study.

In this study, \cite{Dittimi2019}, researchers proposed a mobile-based deep learning application for image classification. However, they used Unity and MATLAB for implementation, whereas we used the Android Studio TensorFlow application development template and Python for model training. This study \cite{Hunt2021} investigated the usability of mobile phones for medical image classification. In this project \cite{Cheung2015,Karargyris2012}, researchers designed an application for skin diseases, with attached hardware for accurate detection. In this paper \cite{Pang2019}, an artificial intelligence diagnostic system on mobile Android terminals for cholelithiasis disease is proposed. This article \cite{Curiel2021} discusses the various challenges that arise when deploying a machine learning model on mobile phones. This work \cite{Johny2021} employs edge computing on a mobile device with an integrated web server to diagnose and forecast metastasis in histopathology pictures.

Among the existing studies \cite{Morikawa2021,Cheung2015,Rajendran2019,Hunt2021,Carroll2017,khan2020,Karargyris2012,Dittimi2019,Straczkiewicz2021,Nwe2020}, researchers have developed a machine-learning pipeline and shed light on the medical and general images on mobile phone applications. However, they did not provide the source code and applications that can be used for result replication.

Many research papers explain image classification on mobile phones, but the following are the reasons for reproducing existing work.

\begin{itemize}

\item The existing papers do not shed light on real-time testing and the concerns that may arise when deploying a machine-learning model on mobile phones. For example, some android mobile phones require models having weights in specific data types (Float32 and Unsigned Int8), and if the model is trained on different data types, then the application crashes.

\item The existing papers do not include source code and documentation, which are essential for reproducing the results and developing an application for some other dataset.

\item The existing papers developed different pipelines for various images like plants, tissues, and X-ray/CT-Scan classification. However, we proposed a general application that works for any classification problem by including and excluding substeps in the pipeline.

\item We compared the model's performance on the computer, mobile phone when pictures were loaded from the camera, and mobile phone in real-time.

\end{itemize}

We believe there must be a generalized application that can capture images in real time and from mobile galleries and classify them into various categories; for that purpose, we used existing templates. Using such a template assists in classification problems involving birds, flowers, plants, objects, tissues, and lung cancer classification. We also analyzed the performance of the same model on real-time and digital images, which showed that the performance of the model was highly degraded in real-time analysis. Finally, we present a method for improving the model's performance on a mobile phone.

Section \ref{section:proposedpipeline} provides the technical context and describes the entire seven-step pipeline process. Section \ref{section:results} demonstrates the implementation of the pipeline on covid-19, plants, and cancer classification as a use case and discusses the model performance and technical considerations. Sections \ref{section:discussion} and \ref{section:conclusion} contain a discussion and conclusion, including a link to all the data and codes to reproduce results.

\section{A pipeline for image classification on mobile phones}
\label{section:proposedpipeline}
\label{sec:previouswork}

\begin{figure}[!ht]
\begin{adjustbox}{width=\textwidth}
\includegraphics[width=14cm,height=18cm]{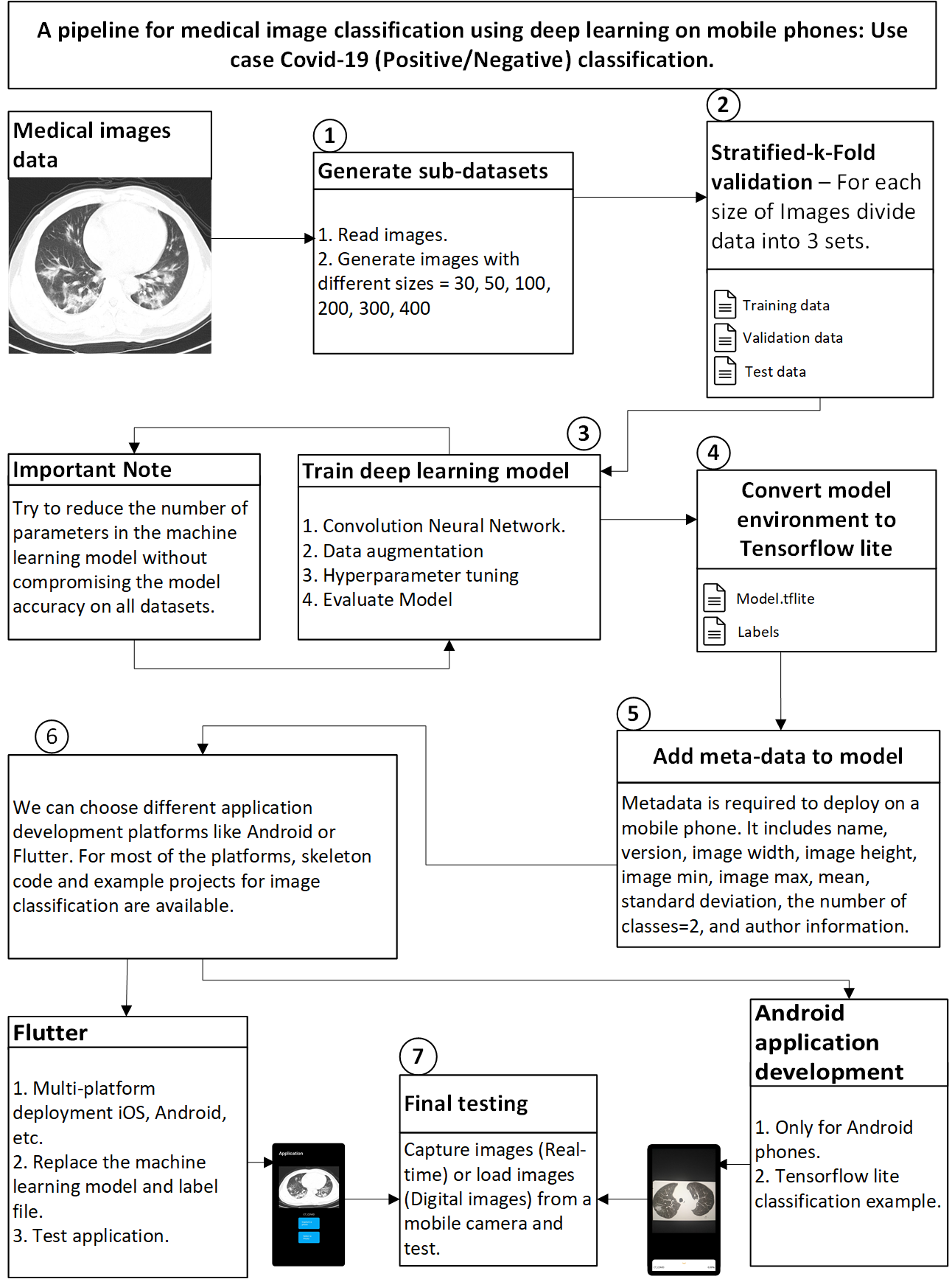}
\end{adjustbox}
\caption{This diagram shows the overall pipeline for implementing deep learning image classification model for mobile phone devices.}
\label{fig1}

\end{figure}

There are already several contexts in which deep learning or other classification models are deployed on mobile phones, including small-scale applications such as emoji selection from text \cite{Bai2019} to larger-scale recommendation systems \cite{9076393,Zhu2017} and face detection from camera images \cite{Choi2011,RosSnchez2020}. Implementing image classification presents many challenges, such as ensuring that the image dimensions are the same for training and testing data. Furthermore, the image background/environment in which the model trained should be the same as in the field; otherwise, the model might be trained to see spurious details in the image background, leading to incorrect results. One must also ensure that the model is implemented efficiently to fit inside mobile memory, often forcing reductions in the model size that can sacrifice accuracy. Finally, if one wants to leverage publicly available medical image data in a mobile context, the details of implementing best practices are unclear. Our answer to all these considerations is a machine-learning pipeline, illustrated in Figure \ref{fig1} and described in this section.

A pipeline is only as good as its data, and starting, one must obtain data suitable for model training and identify the end-user's mobile devices.

As is standard in supervised learning, the training data must be labeled; for medical images, this is typically performed by a domain expert \cite{Hedderich2020}. Many such medical image datasets are available in the public domain \cite{aylwardo70:online,SMIRSICA19:online}, and one should verify labels with the help of a local physician.

\subsection{Step 1: Choose different images sizes and generate sub-datasets}
\label{generatesubdataset}
The first step is preprocessing, which consists of image rescaling, normalization, and image resizing \cite{Nath2014}, and organizing the data appropriately for later analysis. The user selects a small number of different image sizes for testing.
Extra testing in this first step will help avoid excess model parameters, which might crash the mobile application in a later stage.

\begin{figure}[!ht]

\includegraphics[width=8.5cm,height=5cm]{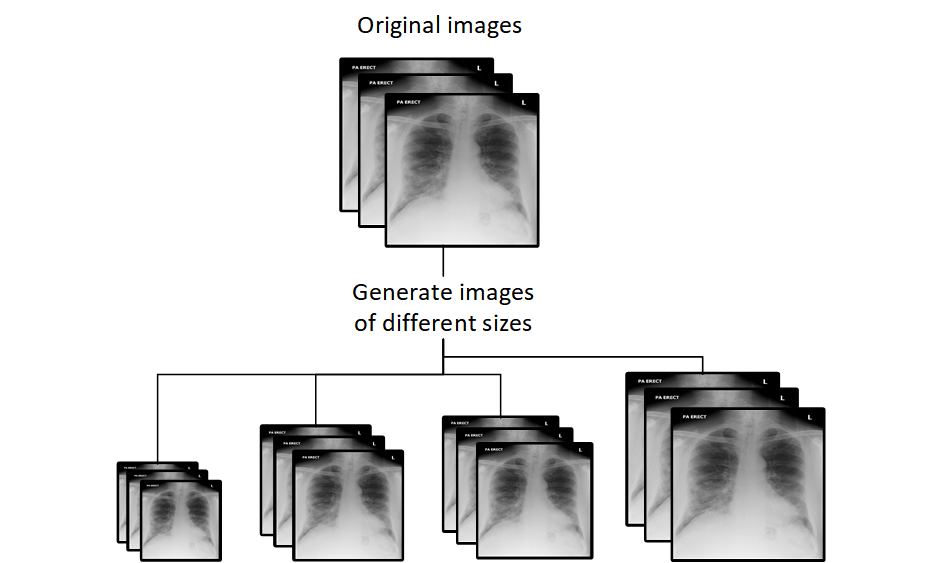}
\caption{
Choose different images sizes and generate sub-datasets.
}
\label{split}
\end{figure}

In practical applications, we recommend choosing 3-5 different sizes ranging from 30 x 30 to 500 x 500 as sufficient, and these choices may depend on the expected aspect ratios of the training data. During the model validation in step 3 \ref{step3}, one of these image sizes will be selected automatically depending on the model performance. Figure \ref{split} shows the subdatasets having different image sizes generated from the original dataset.

\subsection{Step 2: Data splitting for validation}
\label{step2}
It is essential to use stratified-\texttt{k}-fold validation for each image size to avoid over-and under-fitting during the training \cite{Marcot2020}. This ensures that the folds are chosen such that the mean response value is equal across all folds, ultimately decreasing model bias.

Our recommendation is to start with \texttt{k=5}, which repeatedly trains models using 80 percent of the original data and uses the other 20 percent to evaluate model performance.

As is typical with cross-validation, one then systematically trains the model on 4 of the 5 folds and uses the held out to assess the model performance, and the results are averaged to estimate overall model performance.

 

\subsection{Step 3: Model Architecture and Training}
\label{step3}
Image analysis using deep learning methods is a rapidly growing field with many algorithms competing over a wide variety of applications (e.g. LSTM and RCNN) \cite{8987922}. For medical image classification, Convolutional Neural Networks (CNN) are the most popular \cite{Yadav2019}.

Convolution is the process of multiplying pixel values by weights and summing them. The first layer of the CNN frequently detects essential characteristics such as horizontal, vertical, and diagonal edges. The first layer's output is then sent to the second layer, which extracts more complex features like corners and edge combinations. Subsequent layers recognize higher-level characteristics such as objects and faces \cite{10.5555/3203489}. Based on the activation map of the last convolution layer, the terminal layer outputs a series of confidence ratings (numbers ranging from 0 to 1) that indicate how probable the image belongs to a specific class.

Models are ultimately fit in Keras using \texttt{model.fit}. But before training the model, we must address a few key considerations. First, choose the number of parameters or neurons in each layer and the number of layers. If there are too many layers, then there is a possibility that the model may overfit. If there are too few layers, the model may not learn applicable features. Second, reduce the number of parameters (one could imagine a systematic dropout algorithm that trade-off model size and accuracy) to enable execution within a mobile phone's memory \cite{Xia2016}, while also minimizing any associated performance penalties. This is another step that is typically adjusted ``by hand," and as a starting point, we recommend following the steps taken in our use case below in Section \ref{section:results}. Other key considerations before model fitting are included in this subsection and are clearly implemented in the shared code linked below Section \ref{section:results}.

\subsubsection{Data augmentation}
\label{subsubsection:dataaugmention}
Data augmentation is another key step for maintaining model performance in real-world settings. It artificially stimulates and induces noise and other transformations on the training images when fitting the model \cite{Shorten2019,Perez2017TheEO}. In unpredictably noisy applications like ours using potentially multiple mobile phone cameras and hospital settings, such steps are essential.

Parameters for the data augmentation process are controlled via an image train generator and must also be specified. More specifically, when dealing with the medical images captured on mobile phones in real-time, it is essential to consider all options and possibilities with data augmentation, eventually settling on a few different combinations of settings. The key data augmentation parameters in our context are:

\begin{itemize}
\item \texttt{rescale}: Rescaling (normalizing the pixels values in the image) is the default parameter used in image preprocessing for training the model. Original pictures are in RGB format, with pixel values ranging from 0-255. Such numbers would be too large for the model to cope, resulting in an inflating gradient during the backpropagation phase when training the model. So we multiply the data by \texttt{1/255} to change the pixels values between 0-1.
\item \texttt{rotation range, horizontal flip, and vertical flip}: These parameters randomly rotate and flip training data and should be included because mobile photos can be taken in various rotations. The rotated image is appended with pixels that degrade the classification accuracy in the data augmentation phase. For medical images, the horizontal or vertical flip is fine. However, when the image is rotated, we lose a vast amount of information, depending on the rotation range. One vital point to notice here is that rotated images generated by train generators have appended pixels and can degrade performance. In contrast, images captured by rotating phone cameras do contain all the information.

\item \texttt{brightness\_range}: The brightness range in the train generator increases or decreases the image's color brightness to produce multiple images. When the application is deployed in real-time, there is a possibility that the image's brightness captured from a mobile phone is different from the one on which the model is trained. So, this parameter is compulsory in the data augmentation phase to train the model on images of varying brightness.

\item \texttt{zoom range, shear range, width shift range, and height shift range}:
Zoom range randomly zooms inside pictures, shear range applies shearing transformations to image, width, and height shift change image dimension horizontally and vertically \cite{Building55:online}. In a typical image classification task, these parameters can play an essential role in making the model robust. In the case of medical images, it is possible that if these factors are added, the model's performance will suffer, as evidenced by the findings. In medical images, the difference between the positive and the negative case is subtle. For example, in the dog/cat classification problem, we can distinguish them quickly, but in cases/controls, there is just a white pattern in the image. The transformed image produced using these parameters can convert the positive case into negative and vice versa.

Table \ref{tabel2} contains our recommendation of 4 train generators and their respective parameter settings in Keras. Figure \ref{direc} shows the resulting directory structure after following the above steps.

\begin{figure*}[!ht]
 
\includegraphics[width=11cm,height=9cm]{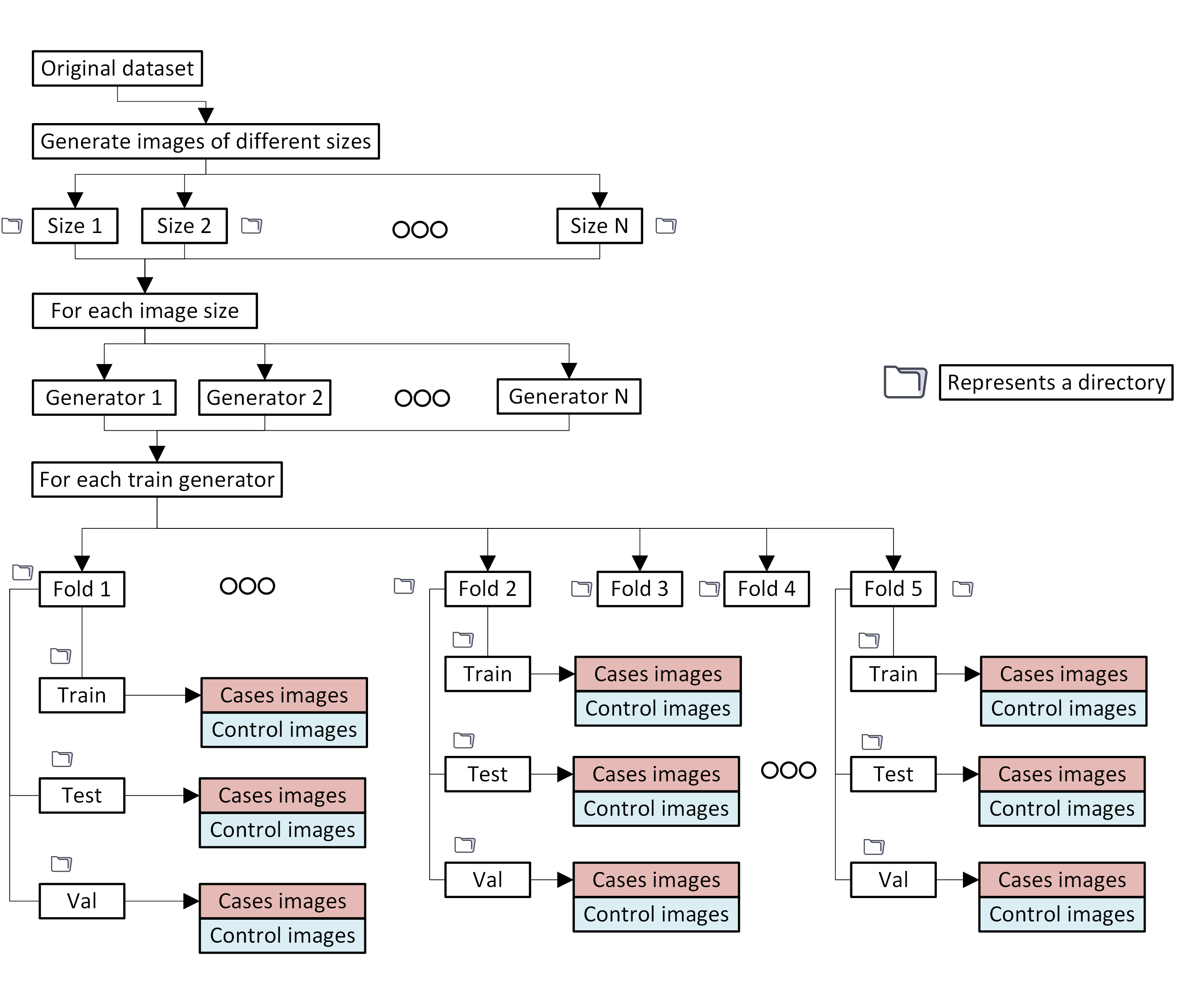} 
\caption{This diagram shows a directory structure for training the model. Generate images with multiple sizes and make a folder for each size. There can be various train generators for each image size, which do not require a separate folder. For each image size, make one folder for each fold. To use data augmentation or train generator, make three folders (train, test, validation) containing sub-folders representing each category.
}
\label{direc}
\end{figure*}

\end{itemize}
\begin{table*}[]
\begin{adjustbox}{width=\columnwidth,center}
\begin{tabular}{|r|l|l|l|l|}
\hline
\textbf{Image Generator Parameters} & \textbf{Generator 1} & \textbf{Generator 2} & \textbf{Generator 3} & \textbf{Generator 4} \\ \hline
\texttt{rescale = 1./255} & \checkmark & \checkmark & \checkmark & \checkmark \\ \hline
\texttt{rotation\_range = 40} & & \checkmark & \checkmark & \checkmark \\ \hline
\texttt{brightness\_range = {[}0.2,1.0{]}} & & \checkmark & \checkmark & \checkmark \\ \hline
\texttt{horizontal\_flip = True} & & \checkmark & \checkmark & \checkmark \\ \hline
\texttt{vertical\_flip = True} & & \checkmark & \checkmark & \checkmark \\ \hline
\texttt{fill\_mode = 'nearest'} & & \checkmark & \checkmark & \checkmark \\ \hline
\texttt{featurewise\_std\_normalization = True} & & & \checkmark & \checkmark \\ \hline
\texttt{featurewise\_center = True} & & & \checkmark & \checkmark \\ \hline
\texttt{zoom\_range = 0.2} & & & & \checkmark \\ \hline
\texttt{shear\_range = 0.2} & & & & \checkmark \\ \hline
\texttt{width\_shift\_range = 0.2} & & & & \checkmark \\ \hline
\texttt{height\_shift\_range = 0.2} & & & & \checkmark \\ \hline
\end{tabular}
\end{adjustbox}
\caption{Recommendations for data augmentation settings in Keras image data processing. The first column shows parameter settings, and \checkmark in subsequent columns denotes inclusion in the 4 recommended image generators.
}
\label{tabel2}
\end{table*}

\subsubsection{Parameter reduction}
\label{subsection:parameterreduction}
Once one identifies the best model by Stratified-\texttt{k}-fold validation, we must reduce parameters to avoid crashing the mobile application. Hyper-parameter tuning does not, but the number of layers and neurons in each layer affects the model's size. Consider the best model from the previous step has two layers with N filters in the first (convolution layer) and M neurons in the second layer (fully connected layer). Increase filters from 1 to N for the first layer, 1 to M (Neurons) for the second layer, calculate model accuracy and the number of parameters in the model. The number of filters or neurons, the size of the filter, dropout, and strong pooling can be used to reduce the model size.

This step is time-consuming, but once the model with fewer parameters is achieved, it can also be deployed on other devices with low computation power like the Raspberry pi. This model contains a compact form of knowledge learned by the model having high parameters. 


 



\subsection{Step 4: Convert model environment to TensorFlow Lite}
\label{step4}
At this step, we have two options: The first is to convert the model to TensorFlow Lite, and the second is to convert the model to TensorFlow Lite and quantize the model (A quantized model executes some or all of the operations on tensors with integers rather than floating-point values) \cite{Deng2019}.

\subsection{Step 5: Specify appropriate metadata}
\label{step5}
Metadata gives information about the model in addition to its fit weights and architecture. It includes the model name, the input size (image size), and the output size (\# of categories), and must be specified. Table \ref{tab2} gives a starting point for metadata settings. After this step, we have two files: the TensorFlow Lite model and a label text file. One should also verify that the order of categories in the label text file matches the model's prediction order.

\begin{table}[!ht]
 
\begin{tabular}{|l|l|}
\hline
name & model's name \\ \hline
version & v1 \\ \hline
image width & 50 \\ \hline
image height & 50 \\ \hline
image min & 0 \\ \hline
image max & 1 \\ \hline
mean & {[}0{]} \\ \hline
std & {[}255{]} \\ \hline
num\_classes & 2 \\ \hline
author & X \\ \hline
\end{tabular}
 
\caption{Model's metadata parameters and dummy values}
\label{tab2}
\end{table}

\subsection{Step 6: Specify appropriate application development platform}
\label{step6}
Different application development platforms can be used, like a Flutter (\texttt{Quantized TensorFlow Lite model}) or Android Studio (\texttt{TensorFlow Lite model}). An already developed application template can build deep learning applications in this stage. Explore several options and decide on the flow of the application. For example, images can be classified in real-time using a camera feed or captured images. This step is related to application development and will not affect the final result.

\subsection{Execution and final considerations}
\label{step7}
At this point, we have a model deployed on potentially multiple mobile phones capable of medical image classification. Select 3 to 5 images from all categories and test the TensorFlow Lite (Quantized/ Unqunatized) model performance on the computer, and that accuracy is the baseline accuracy. For mobile phones, we recommend testing the final model in two ways. The first is to build the application and load those images from the mobile gallery for evaluation (Flutter-based template) using a quantized model. The second option is real-time processing (TensorFlow template) using an unquantized model.

Just like preprocessing is required to train the model on a computer, there is also a preprocessing engine in mobile to preprocess the images, so there can be severe issues when images are tested from the camera feed. The first is the underlying hardware, and the result for the same image using the same model on a computer and mobile phone can yield different results, and there is no solution to mitigate this problem.

\begin{figure*}[!ht]
\begin{adjustbox}{width=\columnwidth,center}
\begin{subfigure}[b]{0.5\linewidth}

\includegraphics[width=0.75\linewidth]{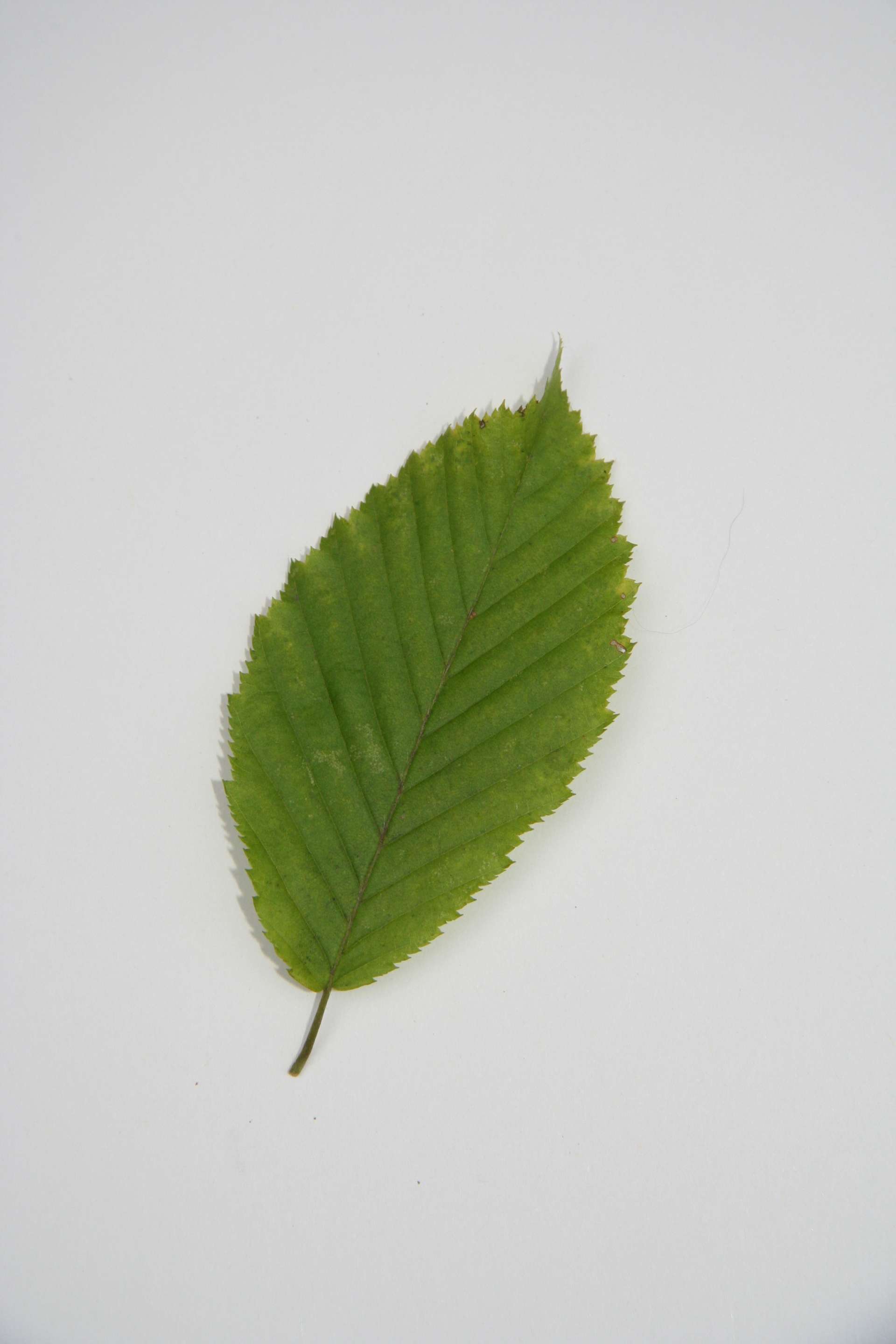} 
\caption{}\label{c1}
\vspace{4ex}
\end{subfigure}
\begin{subfigure}[b]{0.5\linewidth}

\includegraphics[width=0.75\linewidth]{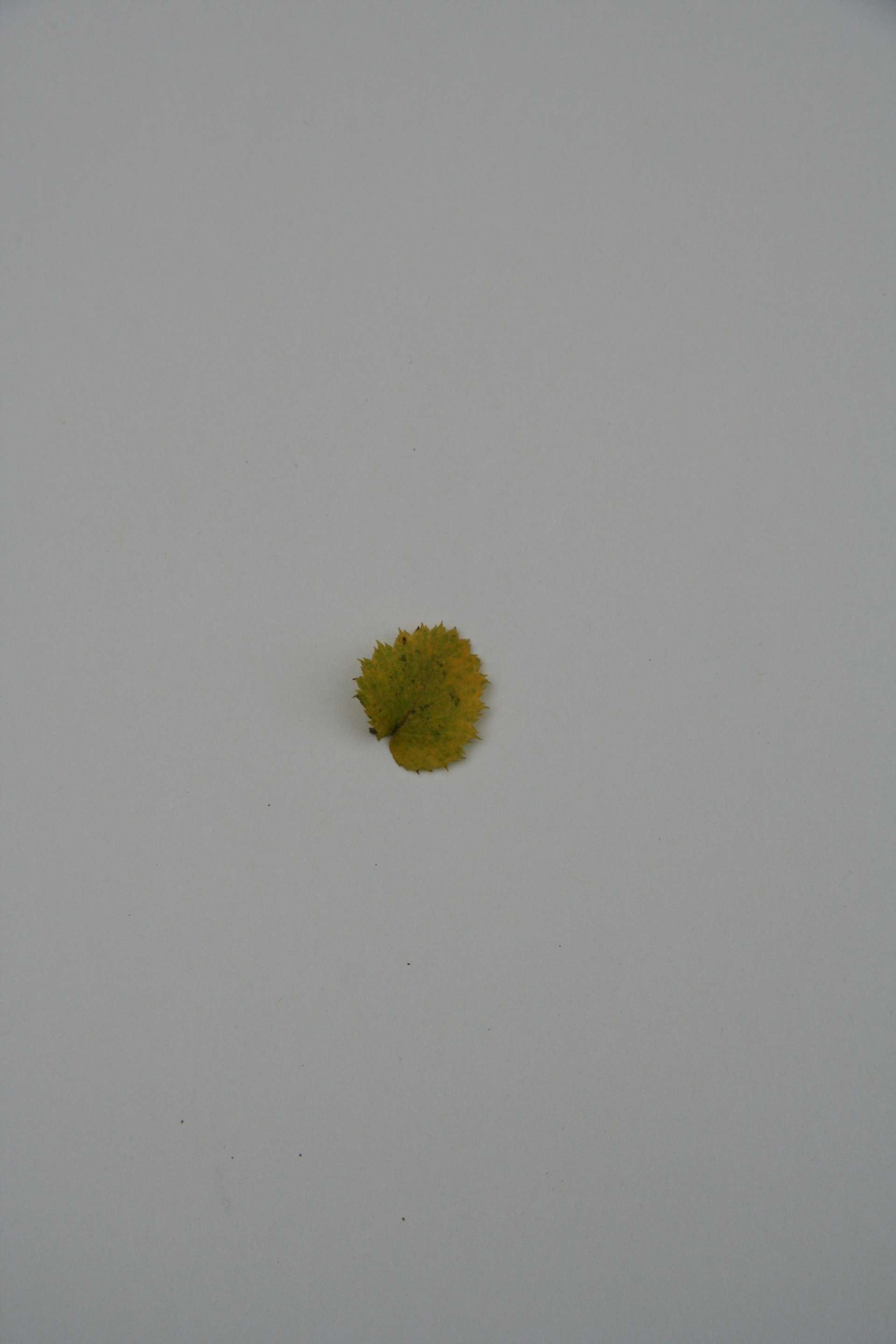}
\caption{}\label{c2}
\vspace{4ex}
\end{subfigure}
\begin{subfigure}[b]{0.5\linewidth}

\includegraphics[width=0.75\linewidth]{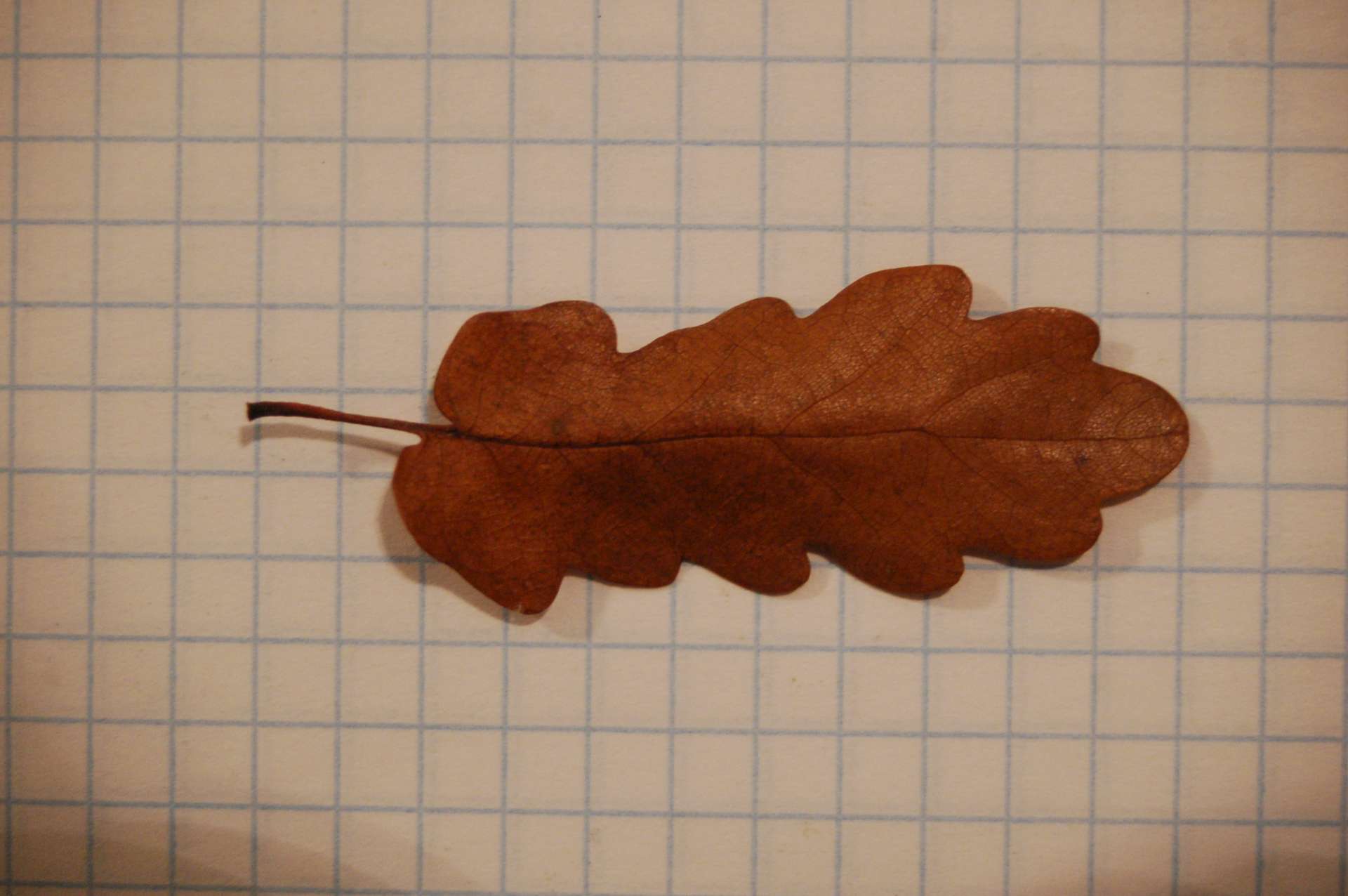}
\caption{}\label{b1}
\end{subfigure}
\begin{subfigure}[b]{0.5\linewidth}

\includegraphics[width=0.75\linewidth]{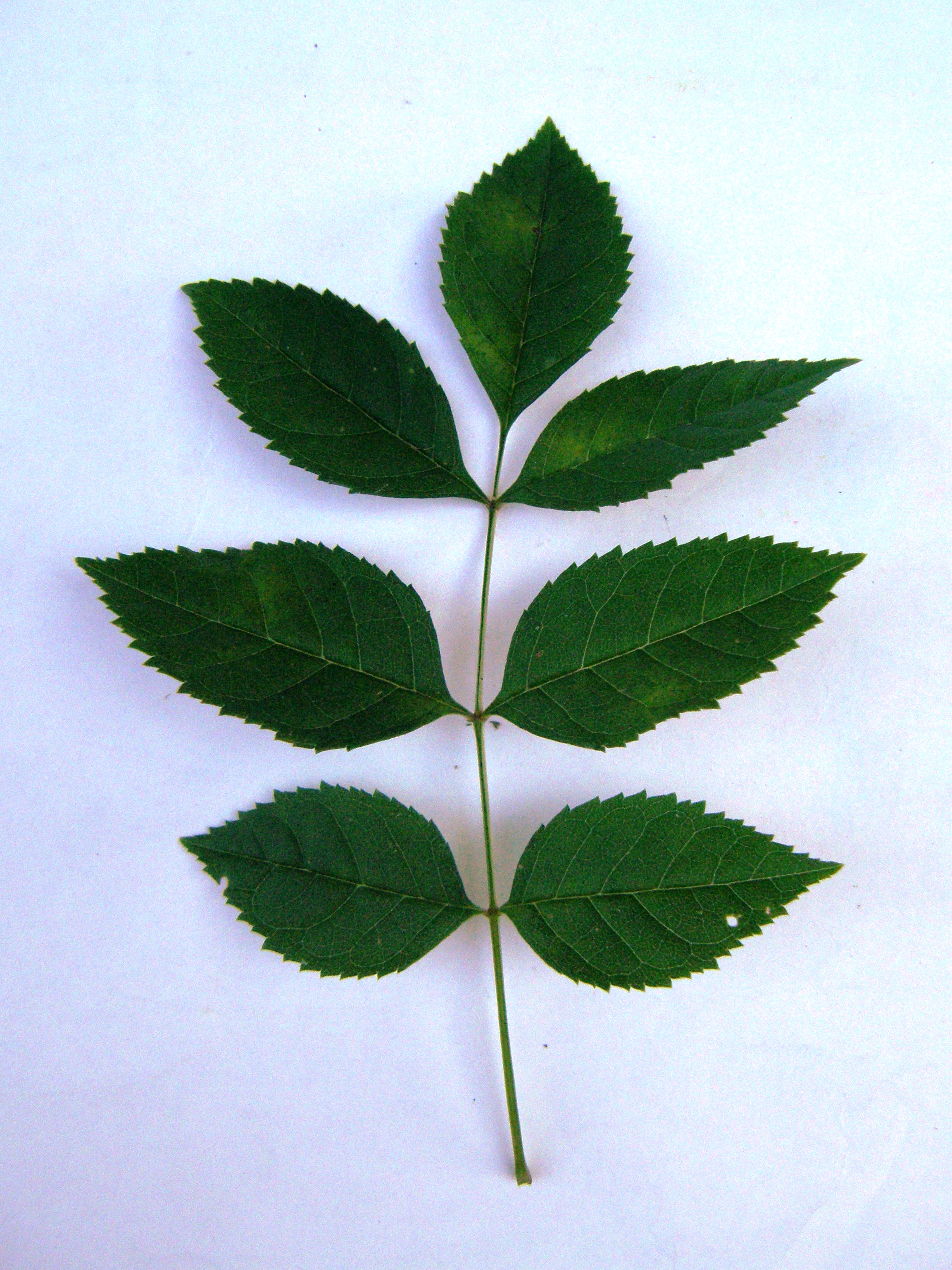}
\caption{}\label{b2}
\end{subfigure}
\end{adjustbox}
\caption{Panels \ref{c1}/\ref{c2}: Inconsistent images of Hornbeam. Panels \ref{b1}/\ref{b2}: Difference of background of various catagories.}
\label{plants}
\end{figure*}

The sizes of the images can vary (See figures \ref{c1} and \ref{c2}), the distance at which the phone should be placed to classify the image can vary, and lastly, the location of images when captured through the phone. If the size of the image varies, then mobile phone distance can be changed such that the frame contains the image. We address each issue separately (See figure \ref{plants}). Lastly, the background of images can vary (See figures \ref{b1} and \ref{b2}).

If the model's accuracy when pictures are loaded from the gallery is low, then the model will not perform when pictures are captured and classified in real-time. So, it is recommended to test the model performance on flutter application before shifting it to real-time.

\begin{figure*}[!ht]

\includegraphics[width=4cm,height=7cm]{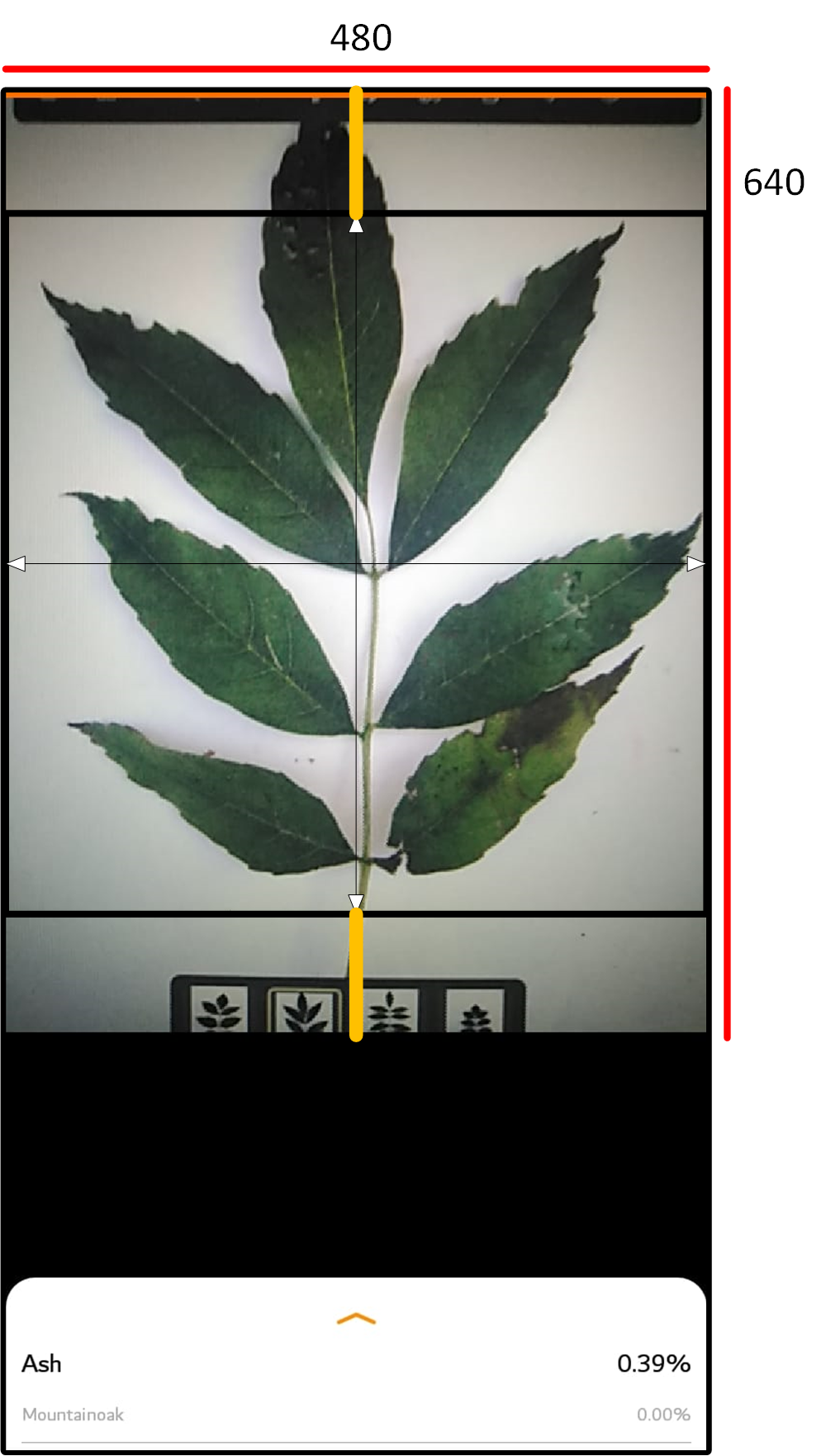}
\caption{ This diagram shows how real-time picture is perceived by mobile phone. The width of the frame in the TensorFlow Lite template is 480 pixels, and the height is 640 pixels. The captured picture is cropped to 480 by 480, as shown by the black box in the middle. The two yellow lines are equal and show that a particular region is ignored.}

\label{tfprocessing}
\end{figure*}

Figure \ref{tfprocessing} shows how the TensorFlow Lite application template perceives an image.

\begin{figure*}[!ht]
\begin{adjustbox}{width=\columnwidth,center}
\includegraphics[width=15cm,height=10cm]{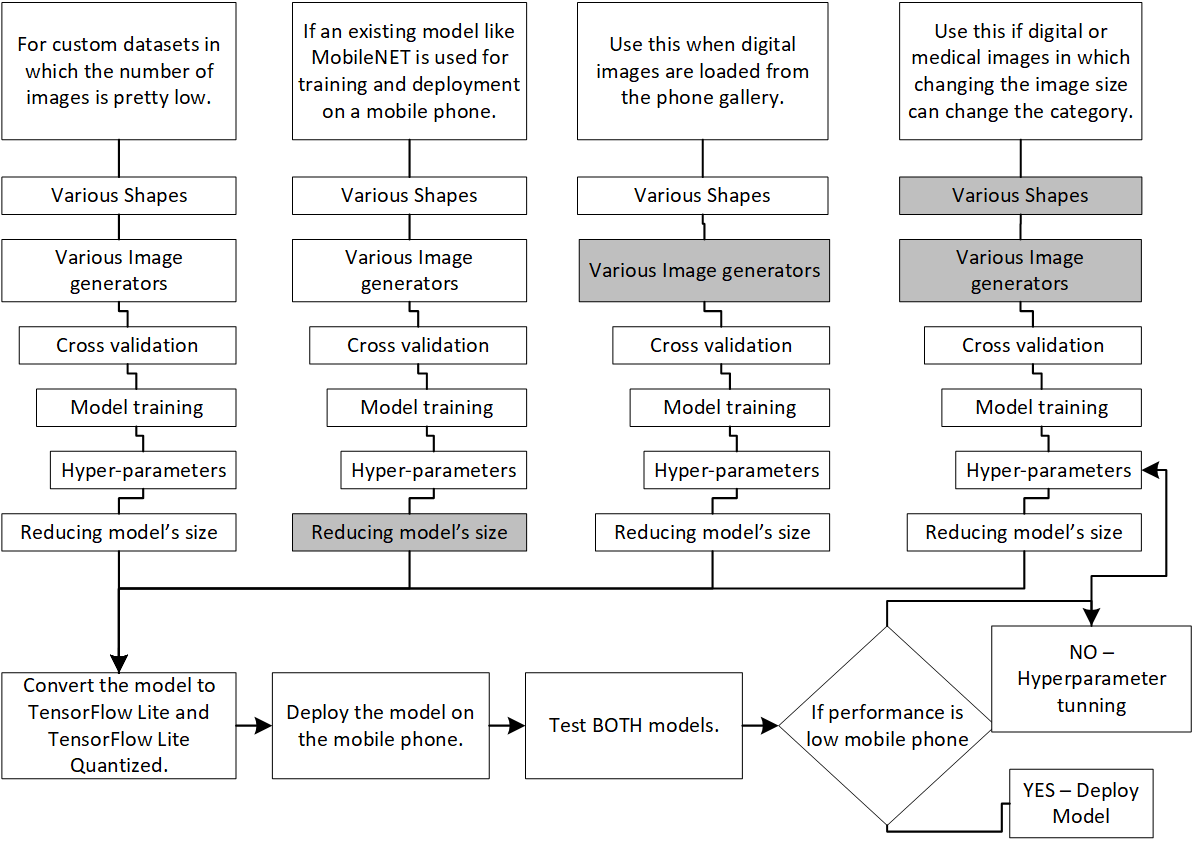}\end{adjustbox}
\caption{This figure shows the four possibilities of using the proposed pipeline depending on the type of images. The grayed boxes skip a particular skip step for a particular images type. For custom datasets, all steps are compulsory. If the existing model is used, there is no need for parameter reduction because the model's size is already optimal. When images are loaded from the gallery image generator, digital images will not increase the accuracy. For digital and medical images like CT scans, the image's size is mostly fixed. If the model's performance when tested on a mobile phone is low, tune model architecture and perform hyper-parameter optimization by mutating batch size, number of epochs, and the number of layers in the model.}

\label{possibilities}
\end{figure*}

If classification accuracy is not sufficiently high, return to step \ref{step3} and repeat the process. To enhance performance, modify the picture size, machine learning model (number of neurons and layers), and hyper-parameters. This final step of testing on additional images is essential for robust performance, and even though the model will run without it, we do not recommend skipping it. Figure \ref{possibilities} shows the possibilities in which the proposed pipeline can be used depending on the type of the dataset.

\section{Use Cases}
\label{section:results}
Medical images have the potential to contribute as a less expensive and more rapid option that can deliver results in minutes instead of days \cite{PlainRad18:online}. The added value is not a replacement for the clinical diagnosis of but to rapidly augment physician information with an uncertain and rapidly evolving virus, thereby improving patient care and outcomes.

The system specifications for the following results are: Intel(R) Core(TM) 7-9750H CPU @ 2.60Hz, 16 GB RAM with NVIDIA GeForce RTX 2060 GPU, running Microsoft Windows 10. The development specifications are Cuda compilation tools release 10.0, V10.0.130, Deep Learning framework Keras 2.4.3, Python 3.6.8, and Tensorflow 2.3.1. The mobile phone specifications are a HUAWEI Y7 Prime 2019, Android version 8.1.0, EMUI version 8.2.0, and Model number DUB-LX1.

\subsection{Dataset 1: Chest X-ray images}
Dataset 1 consisted of images of size (1024,1024,3), which we reduced (Step 1, section \ref{generatesubdataset}) to $(50,50,3)$, $(100,100,3)$, $(200,200,3)$, and $(300,300,3)$. Data splitting for Stratified-5-fold validation (Step 2, section \ref{step2}) resulting in training data (70\%), validation data (10\%), and test data (20\%). The model architecture and training (Step 3, section \ref{step3}) used a CNN with the architecture and hyper-parameters shown in table \ref{arch_1_model_1} and \ref{hyper-parameter}. This data already contained augmented positive/negative X-ray scan images, so we skipped data augmentation (Section \ref{subsubsection:dataaugmention}). Cross-validation, with normal images and augmented images are distributed randomly in the train, validation, and test sets, produced the classification results in Table \ref{table3} with image size of 200 yielding the highest accuracy. Parameter reduction (Section \ref{subsection:parameterreduction}) yielded a model with 4 filters in layer 1 (Convolution layer) and 8 neurons (Fully connected layer) in layer 2, resulting in a reduction in model size from 2,374KB to 24KB. Table \ref{table3} shows that the best performance was found with an image size of 200. Taken together with the results from parameter reduction, the model parameters for metadata were finalized and are shown in Table \ref{tab3}.

\begin{table}[!ht]

\begin{tabular}{|l|l|}
\hline
\multicolumn{2}{|c|}{\textbf{Dataset 1}} \\ \hline
\textbf{Size} & \textbf{Accuracy} \\ \hline
50 & 95.88 (+- 1.80) \\ \hline
100 & 95.77(+- 1.65) \\ \hline
200 & 96.54(+- 1.87) \\ \hline
300 & 59.80(+- 19.41) \\ \hline
\end{tabular}

\caption{
Dataset 1 \cite{dataset1} cross validation accuracy for different image sizes.
}
\label{table3}
\end{table}

\begin{table}[!ht]

\begin{tabular}{|l|l|}
\hline
name & \textbf{Dataset 1 Model} \\ \hline
version & \textbf{v1} \\ \hline
image width & \textbf{200} \\ \hline
image height & \textbf{200} \\ \hline
image min & \textbf{0} \\ \hline
image max & \textbf{1} \\ \hline
mean & \textbf{{[}0{]}} \\ \hline
std & \textbf{{[}255{]}} \\ \hline
num\_classes & \textbf{2} \\ \hline
author & \textbf{X} \\ \hline
\end{tabular}

\caption{Model's metadata parameters for the best model of dataset 1.}
\label{tab3}
\end{table}


This reduced model was converted to TensorFlow lite (Section, \ref{step4}), and the associated code snippets and scripts can be found in this paper's code repository. These steps involved running the TensorFlow lite converter, generating an appropriate label file with classes, and adding the Table \ref{tab3} metadata to the TensorFlow Lite model. This resulted in a model with well-defined input size, range of input values, and output (see Table \ref{tab3}). Then, the TensorFlow Lite Android wrapper code generator was used to create platform-specific wrapper code, which efficiently deploys and executes the model code on the mobile phone \cite{Addingme17:online}. Next, we elected to use the Flutter image classification template (Section \ref{step6}), allowing users to capture an image with their camera and pass it from their phone's image gallery application to the model, which then gives the positive or negative prediction.

\subsection{Dataset 2: Chest CT images}
Dataset 2 \cite{dataset2} contained CT-scan images of size $(1024,1024,3)$, which we reduced (Step 1, section \ref{generatesubdataset}) to $(30,30,3)$, $(50,50,3)$, $(100,100,3)$, $(200,200,3)$, and $(300,300,3)$. The model architecture and training (Step 3, section \ref{step3}) used two CNN models shown in Table \ref{arch_2_model_1} (Model 1) and \ref{arch_2_model_2} (Model 2) with the data augmentation, and cross-validation to train the model. Table \ref{tabel4} shows the result of classification for dataset 2.

\begin{table*}[!ht]

\begin{tabular}{|l|l|l|l|l|}
\hline
\textbf{DataSet 2 / Model 1} & \textbf{Gen1} & \textbf{Gen2} & \textbf{Gen3} & \textbf{Gen4} \\ \hline
\textbf{Shape=30} & 63.11(+- 2.98) & 59.69(+- 6.21) & 61.05(+- 6.48) & 59.20(+- 5.11) \\ \hline
\textbf{Shape=50} & 63.93(+- 4.33) & 57.82(+- 2.84) & 59.60(+- 2.80) & 57.87(+- 6.51) \\ \hline
\textbf{Shape=100} & \textbf{65.82(+- 4.63)} & 62.85(+- 4.13) & 57.65(+- 2.70) & 53.41(+- 6.60) \\ \hline
\textbf{Shape=200} & 59.56(+- 3.58) & 56.23(+- 7.67) & 54.22(+- 5.46) & 53.56(+- 3.58) \\ \hline
\textbf{Shape=300} & 60.16(+- 9.06) & 52.37(+- 6.17) & 56.04(+- 6.30) & 55.08(+- 2.91) \\ \hline
\textbf{DataSet 2 / Model 2} & \textbf{Gen1} & \textbf{Gen2} & \textbf{Gen3} & \textbf{Gen4} \\ \hline
\textbf{Shape=30} & 50.95(+- 6.31) & 54.89(+- 5.13) & 60.68(+- 3.05) & 55.77(+- 2.27) \\ \hline
\textbf{Shape=50} & \textbf{63.98(+- 1.88)} & 60.41(+- 3.96) & 58.87(+- 4.76) & 56.13(+- 5.77) \\ \hline
\textbf{Shape=100} & 59.18(+- 3.80) & 55.30(+- 4.00) & 56.45(+- 3.53) & 55.41(+- 10.60) \\ \hline
\textbf{Shape=200} & 62.94(+- 5.49) & 53.93(+- 4.03) & 56.30(+- 3.41) & 50.39(+- 4.00) \\ \hline
\textbf{Shape=300} & 60.74(+- 6.91) & 50.25(+- 3.97) & 51.92(+- 0.81) & 53.33(+- 5.81) \\ \hline
\end{tabular}

\caption{
We used 5-fold validation, 5 different sizes of input images, and 4 generators. If the accuracy is too low, try multiple models.
}
\label{tabel4}
\end{table*}
The best accuracy for dataset 2 was for model 1 with image input size 100, model 1, and train generator 1. Parameter reduction (sub-section, \ref{subsection:parameterreduction}) reduced the size of the model from 2,374KB to 323KB. Figure \ref{parameters2} (See supplementary material, \ref{supplementarymaterial}) shows the heatmap of accuracies for each combination of filters and neurons in the first layer and the second layer. Convert model environment to TensorFlow lite version (Step 4, section \ref{step4}).


Specify appropriate metadata (Step 5, section \ref{step5}) adds metadata to the TfLite version of the model. The best accuracy for dataset 2 was for model 1, image size 100, so only change the metadata fields mentioned in Table \ref{tab33}.

\begin{table}[!ht]

\begin{tabular}{|l|l|}
\hline
name & \textbf{Dataset 2 Model} \\ \hline
image width & \textbf{100} \\ \hline
image height & \textbf{100} \\ \hline
\end{tabular}

\caption{Model's metadata parameters for the best model of dataset 2.}

\label{tab33}
\end{table}

\begin{figure*}[!ht]
\begin{subfigure}[b]{0.20\linewidth}
\begin{adjustbox}{width=\columnwidth,center}
\includegraphics[width=0.90\linewidth]{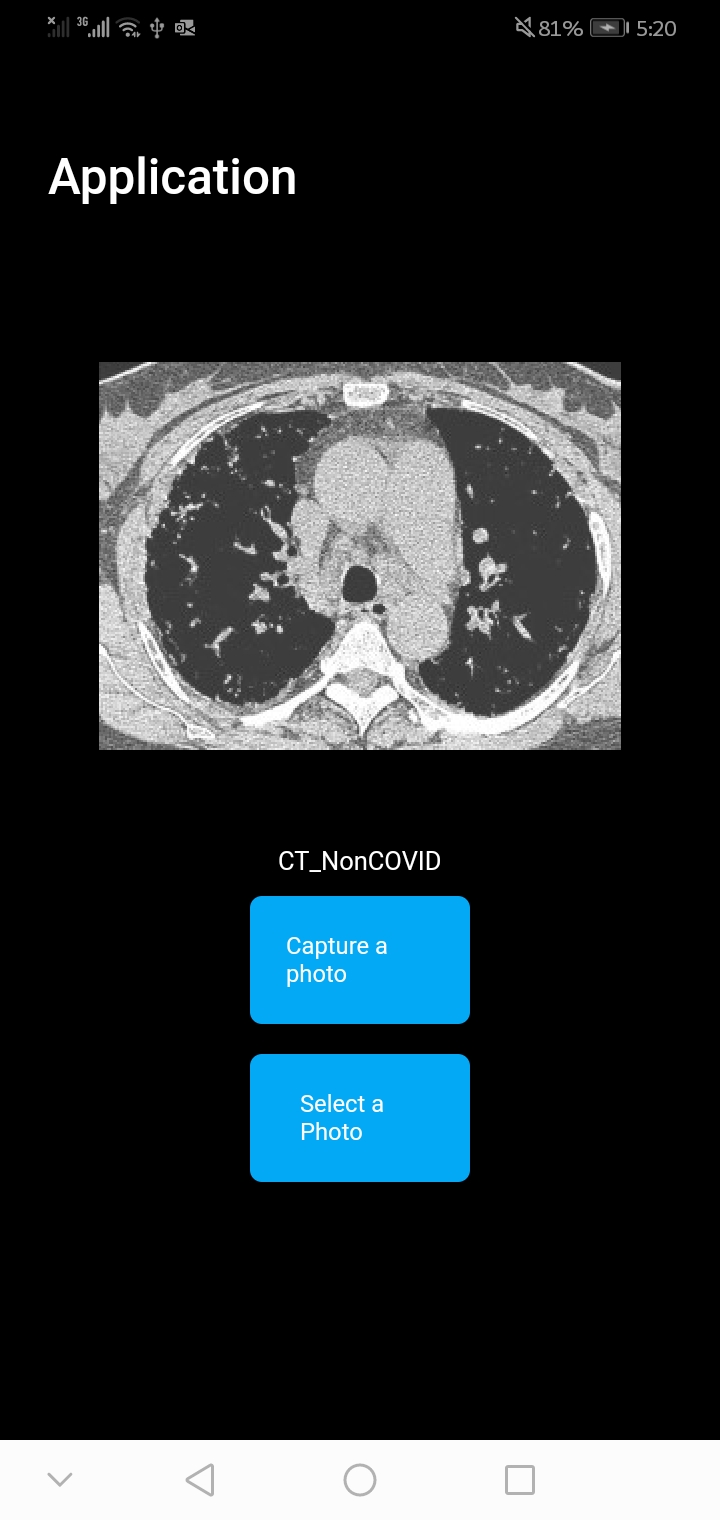}\end{adjustbox}
\caption{}\label{mob3}
\end{subfigure}
\begin{subfigure}[b]{0.20\linewidth}
\begin{adjustbox}{width=\columnwidth,center}
\includegraphics[width=0.90\linewidth]{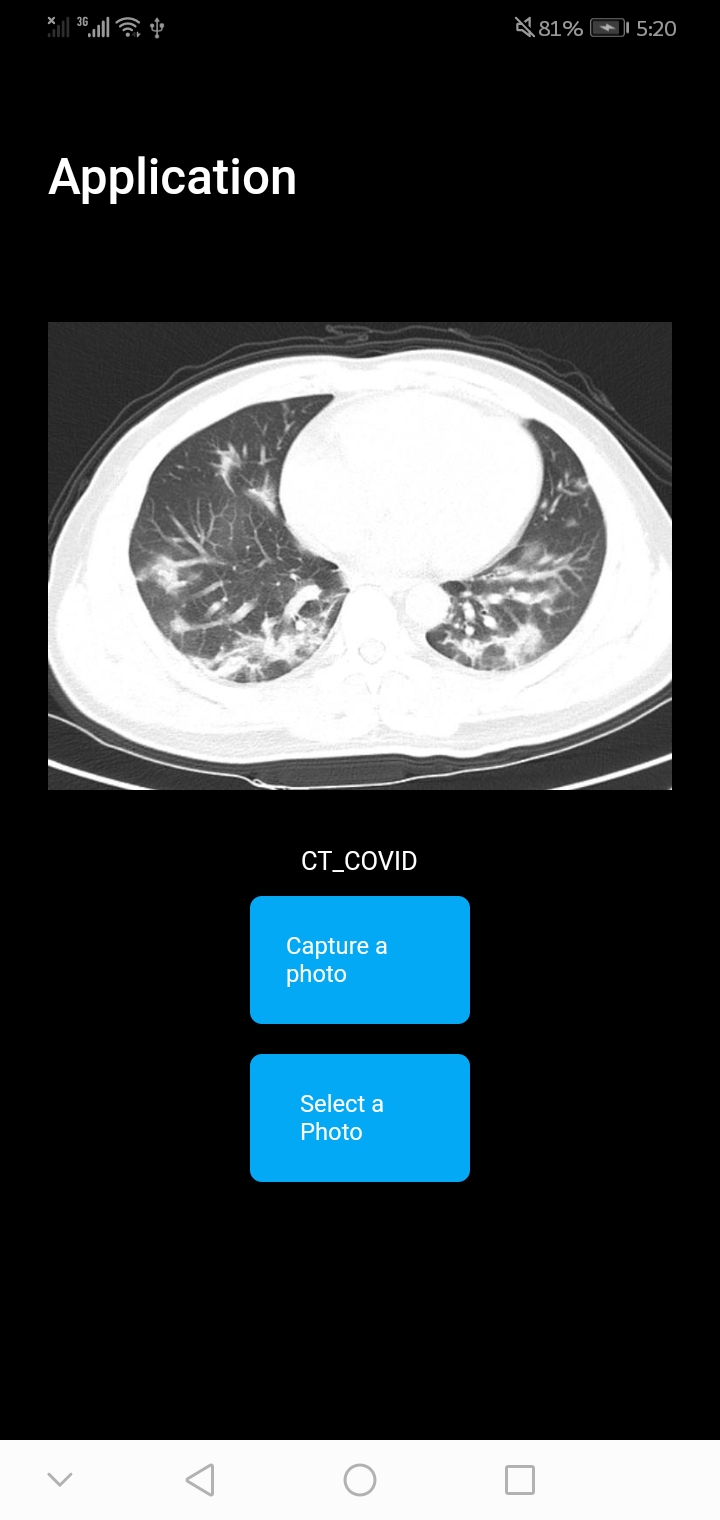}\end{adjustbox}
\caption{}\label{mob4}
\end{subfigure}
\caption{Screenshot of application 1 \ref{mob3} \ref{mob4} for covid detection using CT scan, deployed on android phone. Panels \ref{mob3}/\ref{mob4}: Covid negative/positive Chest CT-scans, respectively.}
\label{appscans}
\end{figure*}

After this step, we have a label file and a model with metadata. Specify appropriate application development (Section, \ref{step6}) used flutter image classification template. Examples of the final application at work can be seen in Figures \ref{mob3} and \ref{mob4}.

In execution and final considerations (Step 7, section \ref{step7}), we considered the first 5 images from both (CT\_COVID and CT\_nonCOVID) categories for the final testing on mobile phone, and those images were not part of model training/testing. We tested the model performance on the mobile phone in real-time, and the final accuracy was 0.6, which means the model cannot be deployed, but the performance increased to 0.80 percent when images were loaded from the gallery (Flutter App).

\subsection{Dataset 3: Leaves}
The dataset \cite{whichfre17:online} consists of five leaves: Ash (24), Beech (28), Hornbeam (30), Mountainoak (20), and Sycamoremaple(20). Dataset is reduced (Step 1, section \ref{generatesubdataset}) to $(50,50,3)$, $(100,100,3)$, $(200,200,3)$, and $(300,300,3)$. The model architecture and training (Step 3, section \ref{step3}) used two CNN models shown in Table \ref{arch_2_model_1} (Model 1) and Table \ref{data_3_model_2} (Model 2) with the data augmentation, and cross-validation to train the model. Table \ref{leaf} shows the result of classification for dataset 3.

\begin{table*}[!ht]

\begin{tabular}{|l|l|l|l|l|}
\hline
\textbf{DataSet 3} & \textbf{50 - Model 1} & \textbf{100 - Model 2} & \textbf{200 - Model 2} & \textbf{300 - Model 2} \\ \hline
\textbf{Gen1} & 87.13 (+- 19.63) & 86.57 (+- 15.74) & 98.04 (+- 2.39) & 97.047 (+- 2.41)\\ \hline
\textbf{Gen2} & 95.03 (+- 4.87) & 91.28 (+- 7.73) & 99.04 (+- 1.90) & 96.04 (+- 1.97) \\ \hline
\textbf{Gen3} & 96.73 (+- 1.64) & 94.09 (+- 7.36) & 96.04 (+- 3.73) & 99.047 (+- 1.90)\\ \hline
\textbf{Gen4} & 92.67 (+- 2.94) & 91.28 (+- 8.18) & 94.14 (+- 5.61) & 91.33 (+- 12.92) \\ \hline
\end{tabular}
\caption{
We used 5-fold validation, 4 different sizes of input images, and 4 generators.
}
\label{leaf}
\end{table*}

Parameter reduction (sub-section, \ref{subsection:parameterreduction}) is skipped. Convert model environment to TensorFlow lite version (Step 4, section \ref{step4}) and produce two files: \texttt{model.tflite} and \texttt{quantizedmodel.tflite}. 


Specify appropriate metadata (Step 5, section \ref{step5}) adds metadata to the TfLite version of the model.

After this step, we have a label file and a model with meta data. \texttt{model.tflite} is deployed on TensorFlow lite template and \texttt{quantizedmodel.tflite} is deployed on Flutter template (Section, \ref{step6}).

In execution and final considerations (Step 7, section \ref{step7}), we considered the first 4 images from all categories for the final testing on mobile phone, and those images were not part of model training/testing. We tested the model performance on the mobile phone when the image size was 50 for generator 3, but the final accuracy was 0.2, which means the model cannot be deployed. At this stage, repeat the process with variation in the model architecture (Step 2, section \ref{step2}) and test the model performance. For shape 224, generator 1, and model 2 (See table \ref{data_3_model_2}), the performance increased to 0.75 percent when images were loaded from the gallery (Flutter App). For shape 224, generator 3, and model 2 (See table \ref{data_3_model_2}), the performance was 0.75 in real-time (TensorFlow App).

\subsection{Dataset 4: Colorectal adenocarcinoma}
This is a set of 7180 image patches (9 different categories) from N=50 patients with colorectal adenocarcinoma \cite{100000hi3:online}. The dataset is challenging to train, test, deploy on the phone, and real testing.

Dataset is reduced (Step 1, section \ref{generatesubdataset}) to $(50,50,3)$, $(100,100,3)$, $(200,200,3)$, and $(300,300,3)$. The model architecture and training (Step 3, section \ref{step3}) used one CNN models shown in Table \ref{data_3_model_2} (Model 1) with the data augmentation, and cross-validation to train the model. Table \ref{cancer} shows the result of classification for dataset 4.

\begin{table*}[!ht]

\begin{tabular}{|l|l|l|l|l|}
\hline
\textbf{DataSet 4} & \textbf{50} & \textbf{100} & \textbf{200} & \textbf{300} \\ \hline
\textbf{Gen1} & 73.19 (+- 6.27) & 71.20 (+- 4.11) & 18.00 (+- 9.54) & 16.80 (+- 8.81)\\ \hline
\textbf{Gen2} & 69.19 (+- 10.24) & \textbf{75.59 (+- 3.20)} & 22.79 (+- 3.91) & 22.39 (+- 2.65) \\ \hline
\textbf{Gen3} & 68.40 (+- 4.96) & 70.00 (+- 4.89) & 17.20 (+- 4.11) & 20.39 (+- 3.44) \\ \hline
\textbf{Gen4} & 70.0 (+- 6.32) & 67.20 (+- 7.44) & 20.79 (+- 3.24) & 22.40 (+- 6.49) \\ \hline
\end{tabular}

\caption{
Cancer classification result.
}
\label{cancer}
\end{table*}
Parameter reduction (sub-section, \ref{subsection:parameterreduction}) is skipped. Convert model environment to TensorFlow lite version (Step 4, section \ref{step4}) and produce two files: \texttt{model.tflite} and \texttt{quantizedmodel.tflite}. 


Specify appropriate metadata (Step 5, section \ref{step5}) adds metadata to the TfLite version of the model.

After this step, we have a label file and a model with meta data. \texttt{model.tflite} is deployed on TensorFlow lite template and \texttt{quantizedmodel.tflite} is deployed on Flutter template (Section, \ref{step6}).

In execution and final considerations (Step 7, section \ref{step7}), we considered the first 4 images from all categories for the final testing on mobile phone, and those images were not part of model training/testing. We tested the model performance on the mobile phone in real-time, and the final accuracy was 0.2, which means the model cannot be deployed. We increased the image size to 200, and the performance increased to 0.56 percent when images were loaded from the gallery (Flutter App). One point to notice here is the test accuracy was 0.99 on the computer.

The following paragraph elaborates the time to execute the pipeline.

For reading images (Step 1, \ref{generatesubdataset}), write a script, and it may take an average time of 20 - 30 minutes, depending on the way the dataset is stored. Below we calculated the time to train the model for images of various sizes for dataset 2, containing about 744 images. Time to train machine learning model (Step 3, \ref{step3}) was 10, 17, 50, 180, and 600 minutes (total time 14 hours) for images having dimensions 30, 50, 100, 200, and 300. The parameter reduction step (Step 4, \ref{step4}) is time-consuming, and for one model having two layers, it took about 1 to 2 days. Converting the model to TensorFlow lite and adding metadata (Step 5, \ref{step5}) takes about 5 minutes. Modifying the image classification template (Step 6, \ref{step6}) for a specific dataset will take about 30 minutes. Building and deploying the application will take about 6 hours for one phone. So the total time for running the pipeline for one dataset is about 3 days.

\section{Discussion}
\label{section:discussion}
This section contains the limitations and future directions of the proposed pipeline. 
There are a few limitations associated with the proposed approach. For example, we considered only android phones (a specific vendor) running a particular version of the Android operating system. There is a high probability that the final classification performance would be identical for phones running the Android operating system due to the same android operating system, but for iPhone or Raspberry Pi, the results may vary. Such applications can also be developed for the iPhone using a different application development framework, one of the future directions for the proposed framework.

\section{Conclusion}
\label{section:conclusion}
This study proposed a pipeline for deploying a deep learning model for medical image classification on mobile phones. The scope of the solution is not only limited to covid-19, but we can also use it for breast cancer or any other medical dataset, making the mobile phone a diagnostic tool for medical images classification. Complex models and other high-end application development skills can also lead to image segmentation. It is essential to highlight the usability and application of the proposed pipeline. Imagine traveling in a deep forest, which has insects and plants that can cause rashes. If someone suffers from skin allergies from a plant, they cannot call a doctor or find any medical assistance. Nevertheless, having a mobile phone application can tell which medicine is appropriate for a particular injury. Such applications can empower doctors in clinical settings where they may require knowledge from other sources to diagnose some diseases better. It will also give access to the public to use that application because there are about 4.3 billion people who use mobile phones.

\backmatter

\section{Supplementary information}

\label{supplementarymaterial}
The documentation associated with the manuscript is available at the following link. \url{https://github.com/MuhammadMuneeb007/A-pipeline-for-image-classification-using-deep-learning-on-mobile-phones}

The code segments associated with the documentation are available at the following link. 
\url{https://muhammadmuneeb007.github.io/A-pipeline-for-image-classification-using-deep-learning-on-mobile-phones/Find\%20a\%20dataset.html#directory-form}

The files, associated applications, and directories are available at the following link.
\url{https://1drv.ms/u/s!AlFVll05llt7gwheV2i4SN3rba13?e=My0QKq}

This section contains the material referenced in the section \ref{section:results}.

\begin{table}[!ht]

\begin{tabular}{|l|l|}
\hline
\multicolumn{2}{|l|}{\textbf{Model 1 architecture for dataset 1}} \\ \hline
Layers & Parameters \\ \hline
Layer 1 - Con2D & \begin{tabular}[c]{@{}l@{}}30 Filters * \\ (kernel size = (3,3))\end{tabular} \\ \hline
Layer 2 - MaxPool2D & (pool size = (2,2)) \\ \hline
Reshape & - \\ \hline
Layer 3 - FullyConnected & (50 Neurons) \\ \hline
Relu & - \\ \hline
Layer 4 - FullyConnected & (2 Neurons) \\ \hline
Softmax & - \\ \hline
\end{tabular}

\caption{Model 1 architecture for dataset 1.}
\label{arch_1_model_1}
\end{table}

\begin{table}[!ht]

\begin{tabular}{|l|l|}
\hline
\multicolumn{2}{|l|}{\textbf{Model’s Hyper-parameters}} \\ \hline
Hyper-parameters & Value \\ \hline
Batch size & 10 \\ \hline
Epochs & 50 \\ \hline
Validation size & 10\% \\ \hline
Optimizer & SGD \\ \hline
Loss & categorical/binary\_crossentropy  \\ \hline
Metrics & Accuracy \\ \hline
\end{tabular}

\caption{Hyper-parameters for all dataset 1 and 2.}
\label{hyper-parameter}
\end{table}

\begin{table}[!ht]

\begin{tabular}{|l|l|}
\hline
\multicolumn{2}{|l|}{\textbf{Model 1 architecture for dataset 2}} \\ \hline
Layers & Parameters \\ \hline
Layer 1 - Con2D & \begin{tabular}[c]{@{}l@{}}32 Filters * \\ (kernel size = (3,3))\end{tabular} \\ \hline
Layer 2 - MaxPool2D & (pool size = (2,2)) \\ \hline
Reshape & - \\ \hline
Layer 3 - FullyConnected & (128 Neurons) \\ \hline
Relu & - \\ \hline
Layer 4 - FullyConnected & (2 Neurons) \\ \hline
Softmax & - \\ \hline
\end{tabular}

\caption{Model 1 architecture for dataset 2.}
\label{arch_2_model_1}
\end{table}

\begin{table}[!ht]

\begin{tabular}{|l|l|}
\hline
\multicolumn{2}{|l|}{\textbf{Model 2 architecture for dataset 3}} \\ \hline
Layers & Parameters \\ \hline
Layer 1 - Con2D & \begin{tabular}[c]{@{}l@{}}32 Filters * \\ (kernel size = (3,3))\end{tabular} \\ \hline
Layer 2 - MaxPool2D & (pool size = (2,2)) \\ \hline
Layer 3 - Con2D & \begin{tabular}[c]{@{}l@{}}32 Filters * \\ (kernel size = (3,3))\end{tabular} \\ \hline
Layer 4 - MaxPool2D & (pool size = (2,2)) \\ \hline
Layer 5 - Con2D & \begin{tabular}[c]{@{}l@{}}32 Filters * \\ (kernel size = (3,3))\end{tabular} \\ \hline
Layer 6 - MaxPool2D & (pool size = (2,2)) \\ \hline
Layer 7 - Con2D & \begin{tabular}[c]{@{}l@{}}64 Filters * \\ (kernel size = (3,3))\end{tabular} \\ \hline
Layer 8 - MaxPool2D & (pool size = (2,2)) \\ \hline
Reshape & - \\ \hline
Layer 9 - FullyConnected & (128 Neurons) \\ \hline
Relu & - \\ \hline
Layer 10 - FullyConnected & (50 Neurons) \\ \hline
Relu & - \\ \hline
Layer 11 - FullyConnected & (20 Neurons) \\ \hline
Relu & - \\ \hline
Layer 12 - FullyConnected & (Categories Neurons) \\ \hline
Softmax & - \\ \hline
\end{tabular}

\caption{Model 2 architecture for dataset 3 and 4. We increased the number of convolutional layers to extract the information because the model \ref{arch_2_model_2} did not work.}
\label{data_3_model_2}
\end{table}

\begin{table}[!ht]

\begin{tabular}{|l|l|}
\hline
\multicolumn{2}{|l|}{\textbf{Model 2 architecture for dataset 2}} \\ \hline
Layers & Parameters \\ \hline
Layer 1 - Con2D & \begin{tabular}[c]{@{}l@{}}32 Filters *\\ (kernel size = (3,3))\end{tabular} \\ \hline
Layer 2 - MaxPool2D & (pool size = (2,2)) \\ \hline
Layer 3 - Con2D & \begin{tabular}[c]{@{}l@{}}64 Filters *\\ (kernel size = (3,3))\end{tabular} \\ \hline
Layer 4 - MaxPool2D & (pool size = (2,2)) \\ \hline
Reshape & - \\ \hline
Layer 5 - FullyConnected & (256 Neurons) \\ \hline
Relu & - \\ \hline
Layer 6 - FullyConnected & (128 Neurons) \\ \hline
Relu & - \\ \hline
Softmax & - \\ \hline
\end{tabular}

\caption{Model 2 architecture for dataset 2.}
\label{arch_2_model_2}
\end{table}

 \begin{figure*}[!ht]
     \begin{adjustbox}{width=\columnwidth,center}
       \includegraphics[width=20cm,height=10cm]{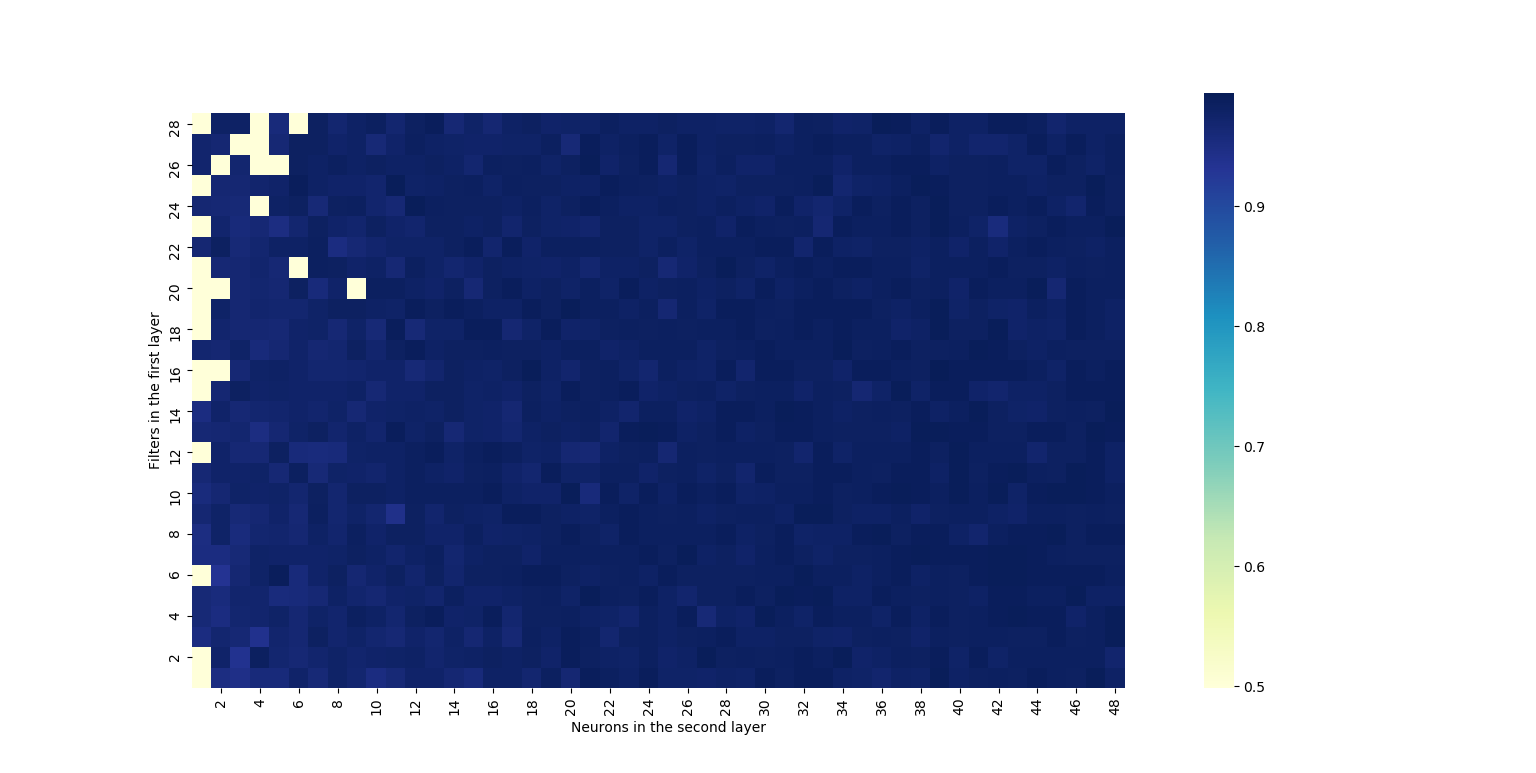}\end{adjustbox}
          \caption{Heatmap of accuracies. Y-axis and X-axis represent the number of neurons in the first layer and the second layer. 
          }  
 
 \label{parameters1}
      
 \end{figure*}

\begin{figure}[!ht]
\begin{adjustbox}{width=\columnwidth,center}
         \includegraphics[width=9cm,height=8cm]{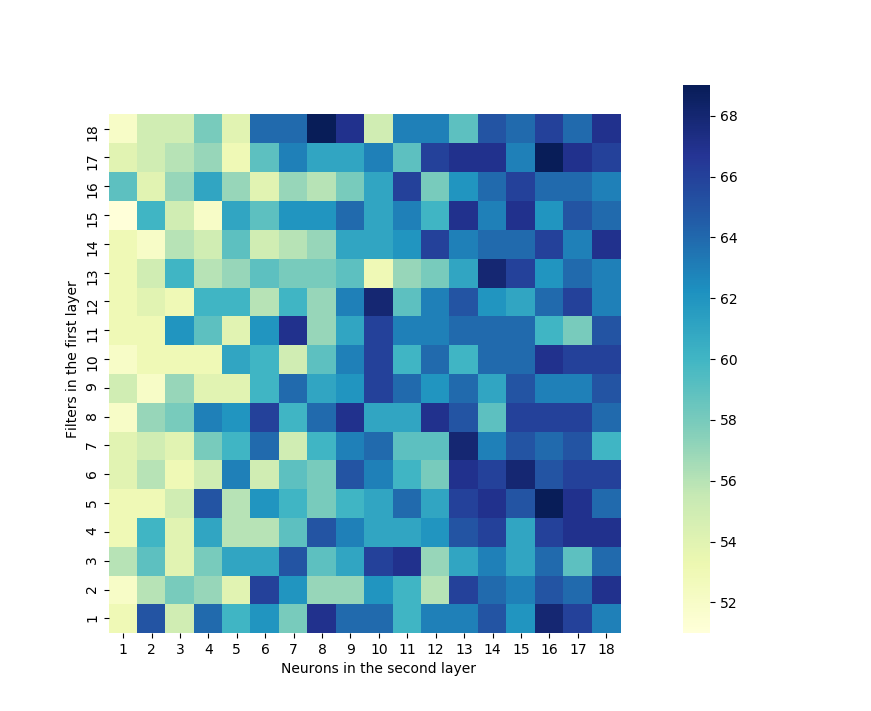}\end{adjustbox}
         
         \caption{Heatmap of accuracies. Y-axis and X-axis represent the number of neurons in the first layer and the second layer. }  
\label{parameters2}
 \end{figure}

\section*{Acknowledgments}
This publication is based upon work supported by the Khalifa University of Science and Technology under Award No. CIRA-2019-050 to SFF.

\noindent

\bibliographystyle{IEEEtran}
\bibliography{bls_final}

\vspace{2cm}

\section*{Authors}
\noindent {\bf Muhammad Muneeb} obtained his M.Sc. in Computer Science from the Khalifa University, Abu Dhabi, UAE. I am currently working as a research associate in the same institute under the supervision of Dr. Samuel. I like to work on inter-discipline problems. Have interests in algorithms, automation, genetics, medical image analysis, and optimization.\\

\noindent {\bf Samuel F. Feng} obtained his PhD in Applied and Computational Mathematics from Princeton University (2012), his MA from Princeton University (2009), and his BA from Rice University (2007). From 2012 to 2014 he was a Ruth L. Kirschstein NRSA Postdoctoral Fellow at the Princeton Neuroscience Institute, working on stochastic models of decision making in psychology and neuroscience. In 2014 he joined the Department of Mathematics at Khalifa University, Abu Dhabi, UAE, where is currently an assistant professor.\\

\noindent {\bf Andreas Henschel} earned his M.Sc. and Ph.D. in Computer Science from the Technical University of Dresden (Germany), in 2002 and 2008, respectively. He joined Masdar Institute as a Post-doctoral researcher in 2009. In 2011, he became Assistant Professor at Masdar Institute, UAE.  He spent one year at the Massachusetts Institute of Technology MIT), USA as a Visiting Scholar.\\

\end{document}